\documentclass[%
 aip,
 amsmath,amssymb,
reprint,%
]{revtex4-1}

\usepackage{graphicx}
\usepackage{xcolor}
\usepackage[english]{babel}
\usepackage{amsmath}
\usepackage[colorlinks=true, allcolors=blue]{hyperref}

\begin{document}

\title{Exploring the potential of ChatGPT for feedback and evaluation in experimental physics}

\affiliation{Facultad de Ciencias, Universidad de la Rep\'ublica, Igua 4225, Montevideo 11400, Uruguay}

\date{\today}

\author{Marcos Abreu$^1$, Alvaro Suárez$^2$, 
Cecilia Stari$^3$ and Arturo C. Marti$^{3,*}$}

\email{marti@fisica.edu.uy}
\affiliation{$^1$
Consejo de Formación en Educación, Administración Nacional de Educación Pública, Florida, Uruguay}

\affiliation{$^2$
Consejo de Formación en Educación, Administración Nacional de
Educación Pública, Montevideo, Uruguay.}

\affiliation{$^3$Instituto de Física, Universidad de la Republica, Montevideo, Uruguay}


\begin{abstract}This study explores how generative artificial intelligence, specifically ChatGPT, can assist in the evaluation of laboratory reports in Experimental Physics. Two interaction modalities were implemented: an automated API-based evaluation and a customized ChatGPT configuration designed to emulate instructor feedback. The analysis focused on two complementary dimensions—formal and structural integrity, and technical accuracy and conceptual depth. Findings indicate that ChatGPT provides consistent feedback on organization, clarity, and adherence to scientific conventions, while its evaluation of technical reasoning and interpretation of experimental data remains less reliable. Each modality exhibited distinctive limitations, particularly in processing graphical and mathematical information. The study contributes to understanding how the use of AI in evaluating laboratory reports can inform feedback practices in experimental physics, highlighting the importance of teacher supervision to ensure the validity of physical reasoning and the accurate interpretation of experimental results.
\end{abstract}

\maketitle
\section{Introduction}

The emergence of Generative Artificial Intelligence (AI), particularly Large Language Models (LLMs), represents a transformative paradigm across multiple fields. Within the Physics Education Research (PER) community, Natural Language Processing (NLP) techniques were already being explored for the large-scale analysis of students’ written responses \cite{bralin2023analysis,Wilson2022Clasification}. More recently, models such as ChatGPT have successfully demonstrated their potential by generating physics-related tasks \cite{liang2023exploring,kieser2023educational,lopez2024challenging,suarez2024chatgpt} and evaluating student work \cite{Kortemeyer2025}, showcasing a remarkable ability to apply domain-specific assessment criteria to complex problems. Despite these advances, the pedagogically effective and responsible integration of these models within the specific domain of physics education remains a significant challenge \cite{Lademann}.

Laboratory reports are multidimensional assessment tools that integrate written explanations, mathematical reasoning, and experimental data. They provide valuable evidence not only for assessing individual student performance but also for revealing how learners engage with authentic scientific practices—designing experiments, interpreting uncertainty, and refining models. When students revise their reports with guided feedback, measurable growth can be observed in their scientific abilities and experimental reasoning \cite{Bugge}. However, evaluating such reports remains a demanding process, with persistent challenges regarding grading consistency and the reliability of feedback in large-enrollment physics courses \cite{passonneau2023ideal}.

While AI is reshaping many dimensions of teaching and learning \cite{zhai2023chatgpt}, its application to the evaluation of laboratory reports is still in its infancy. The complete automation of grading remains unrealistic, given the complexity inherent in validating scientific reasoning and content \cite{Kortemeyer2025}. Nonetheless, recent studies in physics education have begun to explore its potential as a complementary tool for laboratory-based assessment, including the work of Mills et al. \cite{mills2025prompting}, who examined GPT models for providing formative feedback on student reports, and Kilde-Westberg et al. \cite{kilde2025generative}, who tested generative AI as a “lab partner” in experimental activities. These efforts illustrate both the promise and the limitations of AI in supporting authentic scientific reasoning. Effective and responsible integration therefore requires maintaining a balance between automation and expert validation, not only to reduce instructors’ workload but also to preserve the integrity of academic evaluation. Within this perspective, AI opens new possibilities for tracking students’ progress over time, comparing performance across cohorts, and promoting coherence in grading among multiple instructors in large-scale physics courses.

Large language models exhibit a strong dependence on prompt design and training conditions \cite{wei2022chain}. In educational assessment, their capacity to evaluate scientific reasoning and integrity depends not only on the model architecture but also on the way humans interact with it. 
Studies in physics education have shown substantial variability across models and prompting strategies when analyzing students’ laboratory reports and explanations \cite{fussell2025comparing}. 
Developing structured prompts and standardized interaction protocols is therefore essential to achieve consistent and context-sensitive evaluations. 
A key open question is not only whether AI systems can attain a conceptually grounded understanding of students’ reasoning, but also how such depth of analysis can be effectively achieved through appropriate design and interaction strategies.

Within this framework, the present study examines the application of artificial intelligence in the assessment of physics laboratory reports, with the aim of gaining insight into how ChatGPT can contribute to the evaluation process and what limitations arise in this context. The work adopts an exploratory and qualitative approach focused on understanding the potential pedagogical implications of AI-assisted assessment. The guiding research question is: \textit{How can ChatGPT, configured with specific grading rubrics, assist in the evaluation of laboratory reports in Experimental Physics, and what potentials and limitations emerge from its use in this context?} To address this question, the paper is structured as follows. First, we describe the research design and implementation of the AI-assisted assessment protocol. Next, we present the results obtained from its application in an Experimental Physics course. Finally, the discussion and
conclusions examine the strengths, limitations, and potential pedagogical implications of this
exploratory experience.

\section{Research Design and Implementation of AI-Based Assessment Protocols}

To explore the potential of Artificial Intelligence (AI) as a support tool for assessing physics laboratory reports, an evaluation protocol was implemented to analyze the AI’s ability to grade such reports according to criteria established by the teaching staff. The evaluation guide used in this study was developed in 2024 by the course instructors and specifically designed for evaluating reports from the \textit{Simple Pendulum} and \textit{Error and uncertainties} experiments. This guide integrates both formal and conceptual dimensions, ensuring a comprehensive evaluation of students’ performance in the experimental context. The test was conducted at the Faculty of Engineering, Universidad de la República (Uruguay), within the Experimental Physics I course, using laboratory reports produced by students during the 2024 academic year.

A structured protocol was designed to make use of  Artificial Intelligence (AI) as an assistant for report evaluation. This protocol involved the design of a comprehensive prompt that included: (i) assignment of a teaching role to the model, (ii) a checklist of tasks to be performed, (iii) the evaluation guide used by instructors, and (iv) the required output format for responses. The evaluations generated by the AI were then analyzed and compared with the teachers’ assessments of the same laboratory reports, in order to identify strengths, weaknesses, and the degree of agreement between both approaches. A comparative analysis was conducted between two modes of interaction with ChatGPT 5.0, aiming to identify their distinctive characteristics and recognize the inherent limitations of each approach. The protocol was applied to approximately 50 laboratory reports from the 2024 course, providing a representative sample to assess the feasibility of AI-based evaluation in real educational contexts.

The analysis adopted an exploratory and comparative approach structured around two main dimensions. The first, \textit{Formal and Structural Integrity}, addressed the organization of the report according to conventional scientific standards—introduction, materials and methods, results, discussion, and conclusions—as well as textual coherence and adherence to academic conventions, including source citation and graphical presentation. The second, \textit{Technical Precision and Conceptual Depth}, involved the validation of calculations, the consistent and correct use of experimental data, the soundness of scientific reasoning, and the depth of analysis reflected in the discussion and conclusions.

To examine these dimensions, ChatGPT’s performance was evaluated under two interaction modes. In the automated API-based mode, a script submitted the evaluation guide and complete laboratory reports directly to the models, allowing batch processing under controlled and reproducible conditions. This setup reduced the variability inherent to human–AI interaction and enabled a systematic comparison between both modalities. In the customized GPT mode, the model was accessed through its native interface and instructed to emulate a Physics instructor (“Physics Instructor GPT”), enabling context-sensitive and discipline-aligned evaluations.

In the first modality, automated API-based evaluation, a script was developed to submit the evaluation guide and complete laboratory reports to the models via their Application Programming Interfaces (API). This approach allowed for batch processing of reports under controlled and reproducible conditions, minimizing the variability inherent to human–AI interactions and facilitating systematic comparison between both modalities.

In the second modality, customized GPT interaction, each model was accessed directly through its native interface, supplemented with tailored personalization instructions. The AI was configured to emulate the role of a Physics instructor using a custom GPT profile (“Physics Instructor GPT”), which enabled a more nuanced and context-aware evaluation aligned with disciplinary expectations.

\section{Results}

The results obtained from applying the AI-assisted assessment protocol to the laboratory reports from the Experimental Physics course are presented below. The findings are organized according to the two evaluation modalities implemented —assessment via API and assessment with a customized ChatGPT instance— and are analyzed based on the dimensions defined in the evaluation guide: formal and structural integrity, on the one hand, and technical accuracy and conceptual depth, on the other. Finally, the main limitations and discrepancies observed in each modality are discussed, providing a comparative and critical perspective on the performance of AI within this educational context.

\subsection{Evaluation with GPT (API)}

\subsubsection{Formal and Structural Integrity}

The automated assessment performed with GPT revealed a remarkable consistency in addressing the formal aspects of laboratory reports. In several cases, the model highlighted that the ``title was clear, specific and aligned with the purpose of the experiment (determining local $g$ using a simple pendulum'' or that the objectives were “well formulated using appropriate infinitive verbs: to determine g, to apply two measurement methods, to perform statistical analysis.”
The methodology section was also frequently praised for providing a “clear description of the apparatus (plumb bob, string, stand, photosensor, software, stopwatch)” and for including a “schematic representation of the setup (Figure 1)”, strengthening both the structural integrity and replicability of the experiments.
These examples illustrate how GPT was able to accurately identify strengths in writing, organization, and formal presentation, consistently verifying that reports adhered to the expected structure of a standard laboratory report.

\subsubsection{Technical Precision and Conceptual Depth}

When the focus shifted to technical content, the performance of GPT became noticeably more heterogeneous. The evaluated reports included tasks involving mathematical transcription, statistical treatment, and interpretation of results, revealing varying levels of accuracy across these domains. Regarding formula transcription, the model identified several syntax and notation errors—for example, ``Normal distribution incorrectly written (missing the negative sign in the exponent and the $1/(\sigma \sqrt{2\pi})$ factor)'' or ``Equation (4) appears as $\omega = g/l$'' (missing the square root) and (5) does not show the radical in $T = 2\pi/\sqrt{l/g}$.  It also noted that ``several formulas are missing or incorrectly written. The pendulum equation should be $\ddot{\theta} + g/l \sin{\theta} = 0$ and for small oscillations $\ddot{\theta} + g/l \theta = 0$.''

At the statistical level, GPT detected inconsistencies such as the incorrect definition of
standard deviation (not including the square root for example) or the application of data rejection criteria \textit{without sufficient justification}. However, its conceptual interpretation of physical results was often inconsistent: for instance, it pointed out contradictions such as ``it is stated that the values of $g$ are not consistent with each other; however, the confidence intervals overlap,'' or the incorrect conclusion that  ``there is a proportionality between the period and the initial angle.''

However, we must emphasize that many of these corrections proved to be unreliable, as they did not always align with the actual content of the reports. In several instances, GPT flagged non-existent errors or formulas that were correctly expressed by students, revealing an irregular and often inaccurate performance in assessing technical and conceptual aspects.

\subsubsection{System limitations}

A significant portion of the inconsistencies observed can be attributed to intrinsic limitations of the model when processing certain document formats through the API. On multiple occasions, the AI was unable to correctly render images, tables, graphs, or embedded equations. This difficulty often led to partial or incorrect interpretations, which in turn resulted in inaccurate feedback. The system’s responses frequently reflected these interpretive constraints, with comments such as “the cited equations are not included in the text” or “formulas are mentioned but not explicitly written.” In other instances, it identified “errors and ambiguities in mathematical expressions, such as missing roots or exponents in equations.” Additional remarks included “no raw data tables or verifiable units are provided” and “lack of tables and figures/histograms prevents checking titles, axes, or legends.”

To properly interpret these findings, it is crucial to distinguish between genuine issues in the students’ reports and technical constraints in the model’s reading process. In many cases, the AI’s observations did not reflect actual deficiencies in the original content, but rather its difficulty in correctly parsing information. When data, graphs, or equations are not presented in a clear, machine-readable format—i.e., embedded as text rather than as images or screenshots—the API cannot access them, leading the system to produce incomplete or overly critical evaluations, even when the information was present and correctly elaborated.

Similarly, the computational errors flagged by the model did not stem from flaws in the underlying mathematical reasoning, but from incomplete or inaccurate extraction of the relevant information within the reports. In this sense, the limitations observed do not necessarily reflect errors in the students’ work, but rather the system’s restricted ability, when operating through the API, to accurately interpret and process graphical and mathematical elements.

\subsection{Evaluation with custom ChatGPT}

\subsubsection{Formal and Structural Integrity}

The customized ChatGPT model consistently assessed elements such as the report title, formulation of objectives, text organization, figure and table numbering, and the clarity of technical language. It demonstrated an ability to provide balanced feedback—identifying areas for improvement while also recognizing strengths. For instance, it flagged issues such as “The objective should be written using an infinitive verb and should explicitly include the analysis of the time evolution of the period as part of the goal,” or noted deficiencies in figure presentation, suggesting “A schematic with dimensions (sensor position, definition of L to the center of mass) and an explanation of how the oscillation plane was controlled could be included.” At the same time, it also highlighted positive aspects such as “Clear title and well-defined objective: determining g from the pendulum period.”

These examples illustrate the model’s capacity to combine critical observations with acknowledgment of strengths, mirroring the evaluative balance typically expected in academic feedback. The main advantage of this modality lies in the direct interaction with the system, which enables the user to refine the feedback, request clarifications regarding evaluation criteria, and thus create a review process more closely aligned with authentic teaching practice.

\subsubsection{Technical Precision and Conceptual Depth}

In this dimension, greater variability was observed. AI accurately identified key aspects such as the calculation of $g$, the propagation of uncertainties, and the comparison between experimental methods. However, inconsistencies emerged in the interpretation of graphs, tables, and equations. In one report, for example, the model correctly noted: “$g$ is calculated as $4\pi^2 L/T^2$, and the uncertainty propagates assuming $L$ and $T$ are independent variables.” It also demonstrated critical insight when comparing accuracy and precision between methods: “The sensor-based value is precise but biased; the stopwatch-based measurement is consistent due to its greater uncertainty.”

These examples illustrate that model feedback can go beyond purely mechanical assessment, offering conceptually nuanced observations that reflect an understanding of the underlying physics.

\subsubsection{System Limitations}

Several situations were identified in which the model failed to interpret figures, histograms, or equations properly, leading to partial or incomplete evaluations. These issues did not necessarily reflect errors in the students’ work but rather the system’s limitations in recognizing and processing graphical and mathematical elements. In many cases, the students’ reports were complete and accurate, yet the AI’s restricted ability to extract non-textual information resulted in misleading assessments.

Within the automated evaluation process, the model produced remarks such as: “Figures and tables mentioned but not present—for example, references to Figures 1–5 and their appendices, or to histograms and raw-data tables that never appeared”; “Graphs mentioned but the evaluator cannot verify axes or units”; or “Equations cited but not shown; inconsistent equation numbering and absence of raw data.” Such comments illustrate how the system’s interpretation can become partial or ambiguous when reports include visual or symbolic information that it fails to detect correctly. However, when the relevant information—such as images, graphs, tables, or equations—was provided directly through the interface, the system was able to process it effectively and produce accurate evaluations. For example, when a histogram image was submitted, the model responded: “The distribution of periods fits a Gaussian with a mean of 1.96 s and a standard deviation of 0.02 s, confirming the consistency of the experimental data.”

Overall, these cases show that the AI,  due to difficulties in reading and extracting non-textual components, may penalize items even when the information is correctly developed in the report. This pattern was observed repeatedly across reports containing figures, tables, or equations, revealing that the reliability of automated assessment depends strongly on how such elements are embedded and represented within the document.

\section{Discussion and Conclusions}

This exploratory study explored how ChatGPT, configured with specific evaluation guidelines, can assist in the assessment of laboratory reports in Experimental Physics. The findings indicate that the model performs more consistently when addressing formal aspects—such as structure, clarity, and academic writing—than when dealing with technical or conceptual content. These observations directly address the research question, suggesting that ChatGPT may serve as a complementary resource for analyzing students’ reports, provided it is used under teacher supervision to ensure the validity of physical reasoning and the accurate interpretation of experimental results.

Although the AI was not employed to provide direct feedback to students, its consistent performance in reviewing formal aspects of the reports suggests a potential role in supporting instructors with routine checks of structure and presentation. In addition, the aggregated analysis made possible through API-based evaluation points to a potential avenue for identifying recurring difficulties among students—such as issues in data representation, uncertainty analysis, and interpretation of experimental results—thereby informing future instructional adjustments.

When comparing both evaluation modes, the API-based approach proved efficient for large-scale processing but struggled with graphical and mathematical content, whereas the customized ChatGPT interaction generated more contextualized feedback, though with lower reproducibility. These complementary findings reinforce the exploratory character of this study, aimed at understanding the affordances and limitations of AI-based evaluation in experimental physics. Overall, ChatGPT shows promise as a supportive tool for formative assessment, provided it operates under teacher supervision to ensure conceptual validity. Future work should examine the consistency of AI-generated feedback across evaluators and cohorts, and its influence on the development of students’ experimental reasoning.

Beyond these findings, this study highlights the need to interpret AI-generated evaluations not as replacements for human judgment, but as new forms of interaction that can provide feedback practices in the physics laboratory. Encouraging teachers and students to engage critically with such feedback may help integrate these tools into meaningful learning processes, fostering reflection on how evidence, reasoning, and communication interact within experimental work.





\bibliographystyle{iopart-num}
\bibliography{referencias}

\end{document}